\documentclass[twocolumn,twoside,slac_two]{revtex4}
\usepackage{graphicx}
\usepackage{fancyhdr}
\begin{document}

\title{B Spectroscopy at Tevatron}

%

\author{I. Kravchenko}
\affiliation{MIT, 77 Mass Ave, Cambridge, MA 02139}

\begin{abstract}
  Recent results on heavy flavor spectroscopy from the CDF and D0
experiments are reported in this contribution. Using up to 
1~fb$^{-1}$ of accumulated luminosity per experiment, properties 
of $X(3872)$, excited $B^{**}$ states, and the $B_c$ meson are measured. 
Also included are measurements of production rates for ground
state $b$ hadrons in $p\bar{p}$ collisions.
\end{abstract}

\maketitle

\thispagestyle{fancy}


\section{Introduction}

   In this paper we report the most recent results form the 
CDF and D0 experiments on heavy flavor spectroscopy.
We start with the measurements of production fractions
of ground state $b$ hadrons in $p\bar{p}$ collisions
in Sec.~\ref{sec:hadron-production}. Properties of the
X(3872) are discussed in Sec.~\ref{sec:x}. Results on 
the $B_c$ mesons are presented in Sec.~\ref{sec:bc}. Finally,
in the last two chapters the observation and measurement
of masses and widths of $B^{**}$ mesons is presented.


\section{ $\mathbf{b}$ hadron production}
\label{sec:hadron-production}

   The process of transformation of $b$ quarks produced in collisions
into $b$ hadrons is non-perturbative. It can not be easily calculated
and has to be measured experimentally.  In a recent analysis
\cite{cdf-hadron-production}, CDF
measures the probability of a $b$ quark to end up as $B^0$, $B^+$,
$B_s$ or $\Lambda_b$. These probabilities are denoted as $f_d$, $f_u$,
$f_s$ and $f_{\Lambda_b}$, respectively.

\begin{figure*}[t]
\centering
\includegraphics[width=80mm]{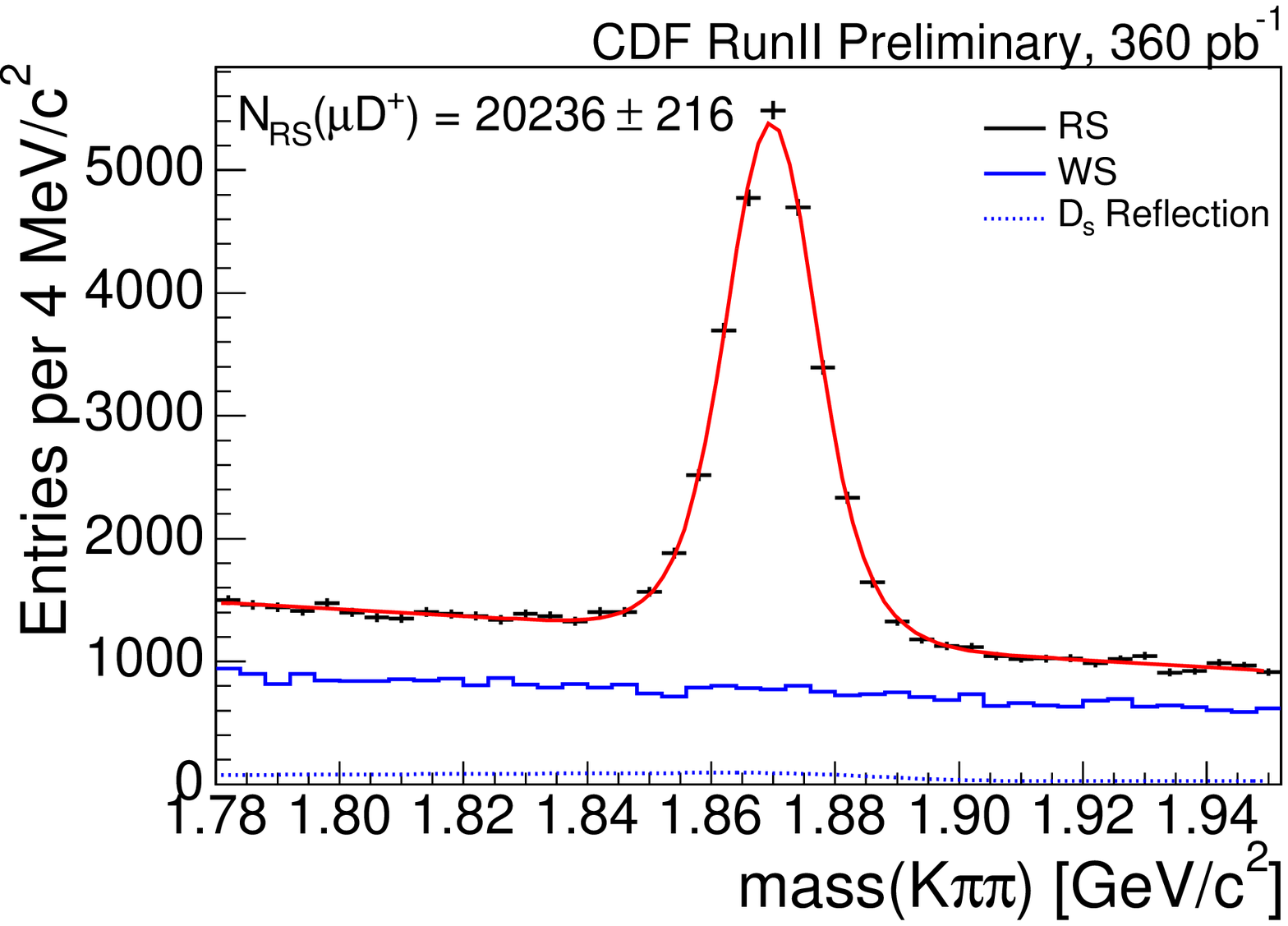}
\includegraphics[width=80mm]{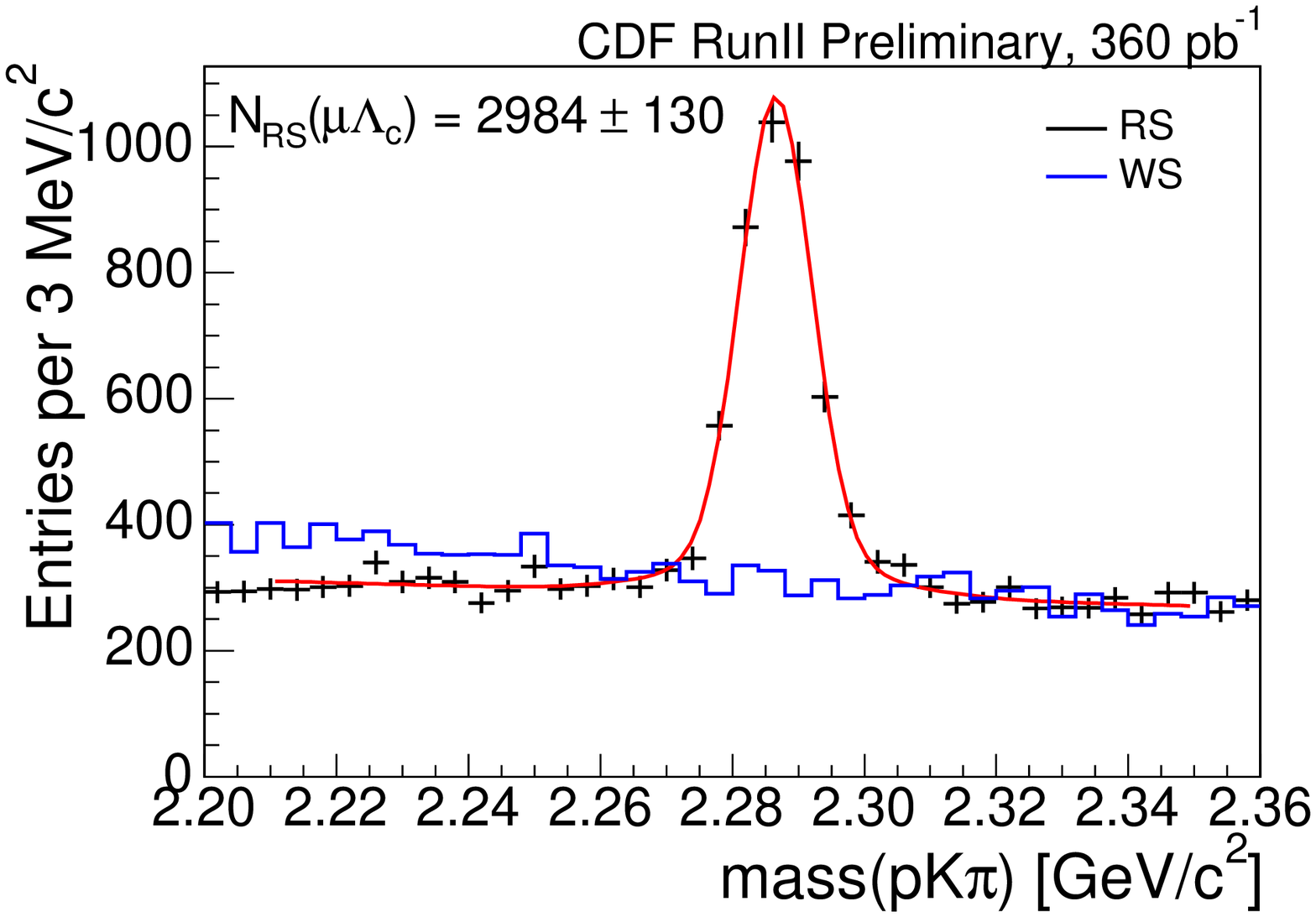}
\caption{Mass spectra for two out of five final states used by CDF
in determination of fragmentation fractions of ground state 
$b$-hadrons. The plot on the left hand side shows the $\mu D^+X$ sample,
and the plot on the right hand side shows the $\mu\Lambda_c X$ sample.} 
\label{fig:b-fragmentation}
\end{figure*}

    Five final states are reconstructed in $~360~pb^{-1}$ of data:
$\ell D^-$, $\ell D^0$, $\ell D^{*-}$, $\ell D_s$ and $\ell\Lambda_c$
where $\ell$ is either an electron or a muon. The invariant mass
spectrum is then fit to obtain the signal yield for
each signature. Two examples of such fits are shown in 
Fig.~\ref{fig:b-fragmentation}.

  One of the difficulties in performing an analysis such as this is
the fact that the final state does not in some cases uniquely identify
the parent ground state $b$ hadron. For example, both $B^0$ and $B^+$
can decay into the $\ell D^0 X$ final state. Therefore, the cross talk
between the channels has to be taken into account. In this analysis
it is assumed that the partial width into the semileptonic channels
is equal for all ground state $b$ hadrons.

 The ratios of the $f$ constants are measured. The ``raw''
ratios containing the branching fractions of the component
decays are found to be the following:

\[
\begin{array}{rl}
 \frac{f_s}{f_u+f_d} & \times Br(D_s^+\to\phi\pi^+) =
 \\
         & (5.76\pm 0.18(stat) ^{+0.45} _{-0.42} (sys))
                                          \times 10^{-3} ~,
\end{array}
\]

\[
\begin{array}{rl}
 \frac{f_{\Lambda_b}}{f_u+f_d}&\times Br(\Lambda_c^+\to pK^-\pi^+) =
 \\
       & (14.1\pm 0.6(stat) ^{+5.3} _{-4.4} (sys))
                                          \times 10^{-3} ~,
\end{array}
\]

\[
\begin{array}{rl}
 \frac{f_{\Lambda_b}}{f_u+f_d} &
               \times Br(\Lambda_b^0\to\ell\nu\Lambda_c) 
               \times Br(\Lambda_c^+\to pK^-\pi^+)  = 
 \\
      & (12.9\pm 0.6(stat)\pm 3.4 (sys))
                                          \times 10^{-4} ~.
\end{array}
\]

From the above, using the best available values for branching 
fractions, CDF obtains:

\[
    \frac{f_u}{f_d} = 1.054\pm 0.018(stat) 
                        ^{+0.025} _{-0.045} (sys)
                        \pm 0.058 (Br) ~,
\]

\[
 \frac{f_s}{f_u + f_d} = 0.160\pm 0.005(stat) 
                        ^{+0.011} _{-0.010} (sys)
                        ^{+0.057} _{-0.034} (Br) ~,
\]

\[
 \frac{f_{\Lambda_b}}{f_u + f_d} = 0.281\pm 0.012(stat) 
                        ^{+0.058} _{-0.056} (sys)
                        ^{+0.128} _{-0.086} (Br) ~.
\]

  The ratios for the $f_u$ and $f_s$ agree well with the values from
the LEP experiments \cite{lep-fragmentation},
however the value of the $f_{\Lambda_b}$ is
currently different from the LEP result by approximately $2\sigma$.


\section{Angular analysis of the $X(3872)$}
\label{sec:x}

  The discovery of a new particle $X(3872)$ was made by BELLE in 2003
\cite{belle-x}. Belle observed this particle in $B$ decays reconstructing the
$J/\psi\pi^+\pi^-$ decay channel of the $X$. CDF and D0 soon confirmed
the discovery. The nature of the $X(3872)$ remained unclear
and is not obvious from the mere observation of a bump in the
$\mu^+\mu^-\pi^+\pi^-$ mass spectrum.  Henceforth, the experiments
capable of producing the $X$, and that includes BaBar in addition to
the three mentioned above, embarked on the program of measuring
properties of the $X$ one by one, as the amount of reconstructed
signal grew over the last years.

   From Tevatron, first, came the mass measurement, then the
production properties including kinematics as well as the prompt
fraction have been measured which suggested
similarity between the $X$ and $\psi(2S)$ production in $p\bar{p}$
collisions \cite{tevatron-X-prop}. 

  One of the best approaches to figuring out the nature of
the $X$ is measuring its quantum numbers. The values of 
$J^{PC}$ are the target of the latest analysis from CDF 
\cite{X3872}. 

%

\begin{figure*}[t]
\centering
\includegraphics[width=120mm]{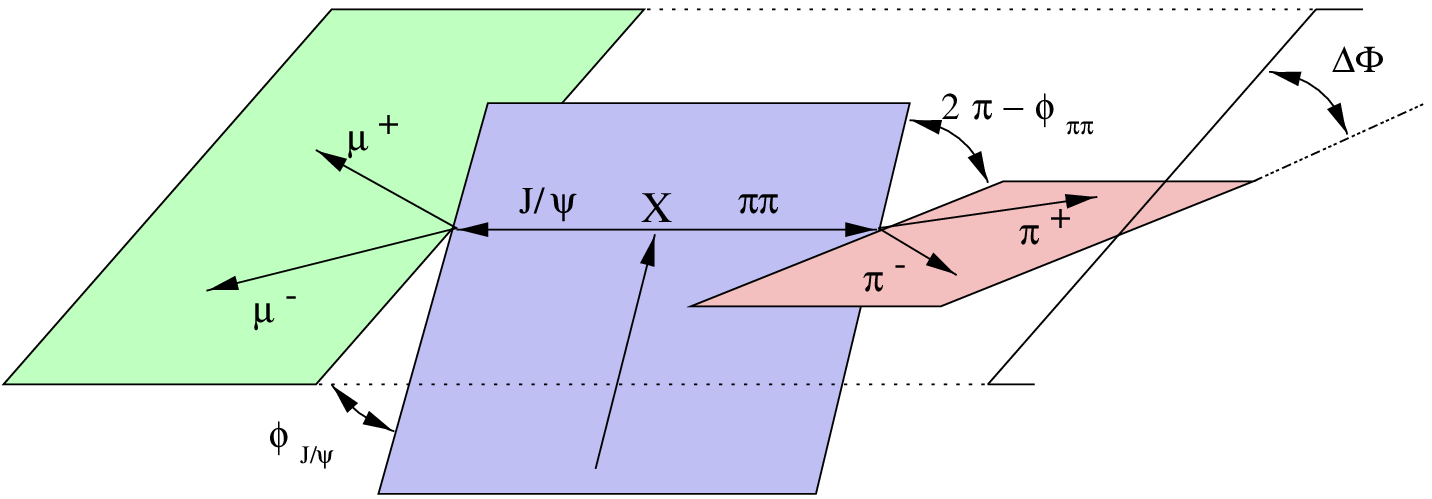} \\
\vspace{1.5cm}
\includegraphics[width=90mm]{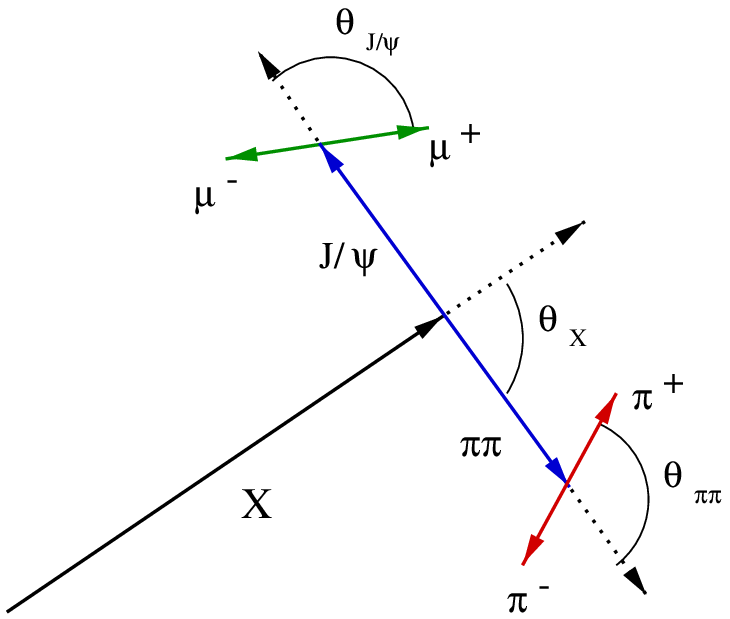}
\caption{
Definition of $\phi$ and $\theta$ angles for the analysis of angular
decay properties of the X(3872).
} 
\label{fig:x-angle-defs}
\end{figure*}

  The $J^{PC}$ is investigated by analyzing angular distributions of
the decay. It turns out that the most sensitive angles to $J^{PC}$ values
are the two helicity angles $\theta_{\pi\pi}$,
$\theta_{J/\psi}$ and the angle $\Delta\Phi$ between the decay planes
of the $\mu\mu$ and the $\pi\pi$ systems.
 These angles are illustrated
in Fig.~\ref{fig:x-angle-defs}. 

\begin{figure*}[t]
\centering
\includegraphics[width=135mm]{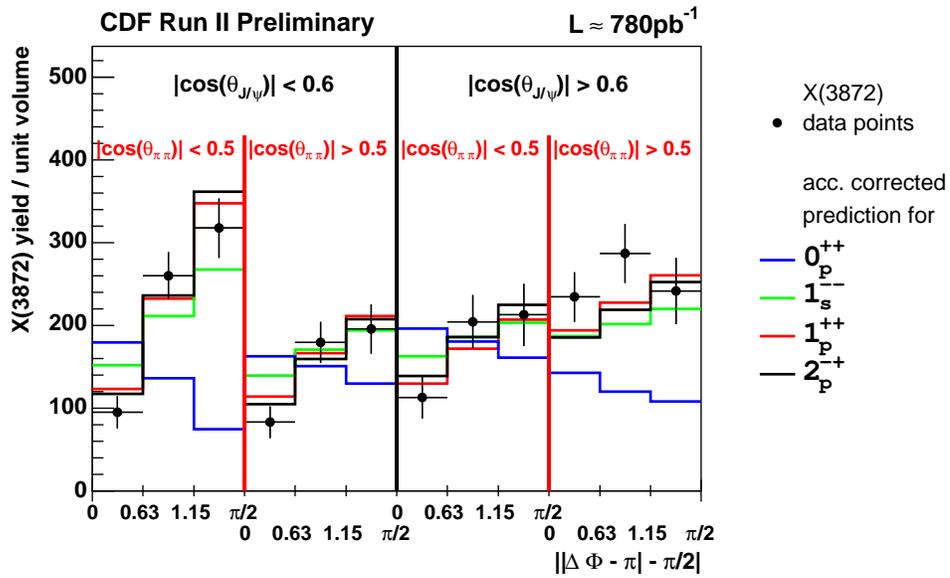}
\caption{
The number of the X(3872) signal candidates as a function
of the two helicity angles and the angle between the decay
planes of the $\pi\pi$ and $\mu\mu$ systems.
} 
\label{fig:X_result}
\end{figure*}

In approximately $~780~pb^{-1}$ of data approximately 3000 of the $X$
particles are found.  In order to reconstruct the angular
distributions for the signal candidates, all reconstructed candidates
are split into bins in three dimensions: three bins in $\Delta\Phi$,
two bins in $\theta_{\pi\pi}$ and two bins in $\theta_{J/\psi}$ giving
total of 3$\times$2$\times$2$=$12 bins.  In each of these subsamples
the invariant mass spectrum of the candidates is fit, and the signal
yield is measured. The result for all 12 bins is drawn in
Fig.~\ref{fig:X_result}.

   The expected shape for the distribution is derived from theory. The
amplitude is factorized according to the isobar model into the three
factors for that describe the $X$, the $J/\psi$ and the $\pi\pi$
systems as well as the two propagators for the $J/\psi$ and $\pi\pi$. 
The angular momentum values $L=0$ and $L=1$ are
considered for the $\pi\pi$ system. The correction to the shapes due
to detector acceptance effects are applied.  The shapes are derived
for a variety of $J^{PC}$ models.

  For each model a $\chi^2$-based probability of matching the 12 data
points is calculated. The list of models and corresponding
probabilities are found in Table~\ref{tab:x-table}. 
The three best and the one least
probable hypothesis are plotted over the data points in
Fig.~\ref{fig:X_result}.  It
is immediately obvious that there are two possible combinations that
fit the data quite well: $J^{PC}=1^{++}$ and $2^{-+}$. All others
are excluded at $>3\sigma$ level.

   As the angular analysis at CDF can not distinguish between
the two possibilities for the $J^{PC}$, one may consider 
examining the $\pi\pi$ invariant mass distribution. The shape
of this distribution is different for the $L=0$ case that corresponds
to the $1^{++}$ and the $L=1$ case corresponding to the $2^{-+}$.
This has been attempted by Belle \cite{belle-pipi} and the conclusion
was, using a simple Breit-Wigner model, favoring $1^{++}$. 
Later, CDF released an analysis of the dipion mass spectrum 
\cite{cdf-x-pipi}
using more sophisticated and generalized approach concluding that
both $L=0$ and $L=1$ are possible. 
And thus, CDF results from both angular and
dipion mass distributions allow either $1^{++}$ or
$2^{-+}$ assignments for the $X(3872)$.

\begin{table}[h] 
\caption{The list of hypotheses considered in the angular
analysis of the X(3872) candidates as well as 
$\chi^2$/probabilities from fits of the data.
}
\begin{center}
\begin{tabular}{|c||c|c|}
\hline
 hypothesis &  3D $\chi^2$ / 11 d.o.f. & $\chi^2$ prob.\\
\hline
 $1^{++}$  &   13.2         &   27.8\% \\
\hline
 $2^{-+}$  &   13.6         &   25.8\% \\
\hline
 $1^{--}$  &   35.1         &   0.02\% \\
\hline
 $2^{+-}$  &   38.9         &   5.5$\cdot10^{-5}$     \\
\hline
 $1^{+-}$  &   39.8         &   3.8$\cdot10^{-5}$     \\
\hline
 $2^{--}$  &   39.8         &   3.8$\cdot10^{-5}$     \\
\hline
 $3^{+-}$  &   39.8         &   3.8$\cdot10^{-5}$     \\
\hline
 $3^{--}$  &   41.0         &   2.4$\cdot10^{-5}$     \\
\hline
 $2^{++}$  &   43.0         &   1.1$\cdot10^{-5}$     \\
\hline
 $1^{-+}$  &   45.4         &   4.1$\cdot10^{-6}$     \\
\hline
 $0^{-+}$  &  103.6         &   3.5$\cdot10^{-17}$    \\
\hline
 $0^{+-}$  &  129.2         &   $\le$1$\cdot10^{-20}$ \\
\hline
 $0^{++}$  &  163.1         &   $\le$1$\cdot10^{-20}$ \\
\hline
\end{tabular}
\end{center}
\label{tab:x-table}
\end{table}


\section{Properties of the $B_c$}
\label{sec:bc}

  Although discovered in 198 by CDF \cite{bc-discovery}, 
the properties of the $B_c$ remain poorly measured due to 
small samples of candidates available until recently.
In Run II, CDF at D0 experiments have finally accumulated
enough data to study the $B_c$ in greater detail. Being the
last discovered ground state $B$ meson and the only 
meson with two heavy quarks of different flavor, the $B_c$
is a great laboratory for potential models, HQET and lattice
QCD. Its mass, lifetime, decay properties and production
are all of interest as many precise predictions have been
made by theorists.

  At the Tevatron, the $B_c$ is reconstructed in several decay channels
containing a $J/\psi$ meson. It is seen in the semileptonic modes
$B_c\to J/\psi e \nu X$ and $B_c\to J/\psi \mu \nu X$ by CDF, and in
$B_c\to J/\psi e\mu \nu X$ by D0. The signal significance in all cases
is over $5\sigma$ and the number of signal events (100$+$) is
sufficient for a proper decay time measurement. D0 uses $J/\psi \mu$
while CDF uses $J/\psi e$ events. The measured values for the proper
decay time are the following \cite{cdf-bc-tau,d0-bc-tau}:

\[CDF: \tau_{B_c} = 0.474 ^{+0.073} _{-0.066} \pm 0.033 ps ~, \]

\[D0:  \tau_{B_c} = 0.448 ^{+0.123} _{-0.096} \pm 0.121 ps ~. \]

Note, that only a fraction of available data is used by both
experiments (CDF analyzed $360~pb^{-1}$ and D0 $210~pb^{-1}$), so
significant improvements of the measurements are expected
in near future.

  The measured value agrees well with the theoretical prediction 
of $0.55\pm0.15$ found in \cite{bc-theory}.
  
  In addition, CDF measures the $B_c$ mass with high precision 
\cite{cdf-bc-mass}
analyzing
a sample of fully reconstructed $B_c\to J/\psi \pi$ decays. In 
$~0.8 fb^{-1}$ of data a clear peak with $>5\sigma$ significance
is seen in the invariant mass spectrum of $\mu\mu\pi$ candidates.
A fit to this spectrum yields the mass measurement:
 
\[ M(B_c) = 6275.2\pm 4.3\pm 2.3 MeV/c^2 ~. \]

\noindent
which agrees moderately well with predictions from lattice QCD
$M(B_c)_{LAT} = 6304\pm 12 ^{+18}_{0} MeV/c^2$ \cite{bc-theory}.


\section{Measurements of $B^{**}$ narrow states}
\label{sec:bstst}

\begin{figure*}[t]
\centering
\includegraphics[width=100mm]{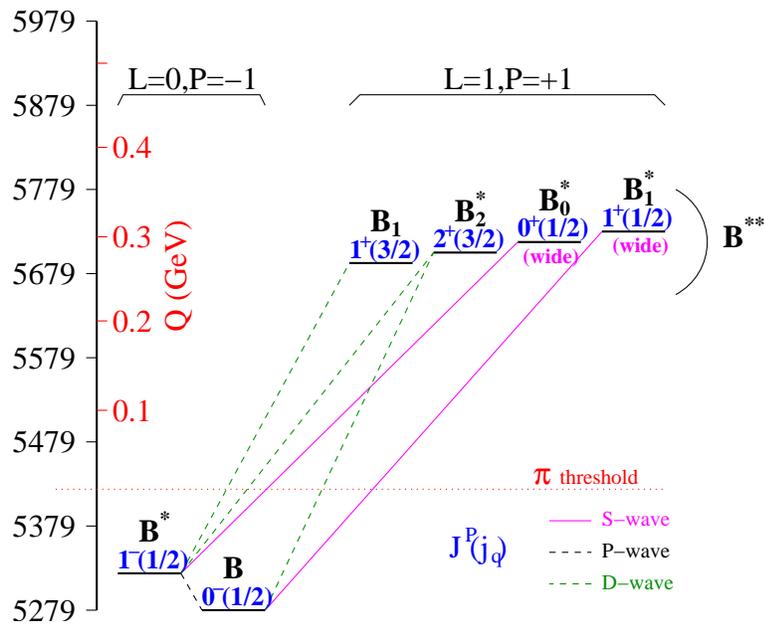}
\caption{States of a $b\bar{q}$ system. The $B$, $B^*$ and
$B^{**}$ states are shown as well as the allowed transitions.
}
\label{fig:bstst-diagram}
\end{figure*}

  The spectroscopy of the $b\bar{q}$ system, where $q$ is either $u$ or $d$
quark, is well understood theoretically. The HQET describes a
heavy-light state and predicts that there are four P-wave states,
collectively called $B^{**}$ or $B_J$, see Fig.~\ref{fig:bstst-diagram}. 
It is expected
that two of them, $B^*_0$ and $B^*_1$ are wide states as they decay
via S-wave. On the other hand, the remaining to states are narrow
because they decay via D-wave. The quantitative understanding is not
nearly as good. Few experimental data are available on $B^{**}$
properties. However, since recently we are starting to see progress in
this area.

  Both CDF and D0 seek to observe and measure the two of
the $B^{**}$ that have a narrow width, expected to be
$O(10) MeV/c^2$ from theory. The other two P-wave states,
the broad ones, are ignored as they are so wide that
distinguishing them from combinatorial background is
nearly impossible with the available data. $B^0_1$ decays 
only to $B^{*+}\pi^-$ while $B^{*0}_2$ can decay to 
either $B^{*+}\pi^-$ or the ground state $B^+\pi^-$.

\begin{figure*}[t]
\centering
\includegraphics[width=135mm]{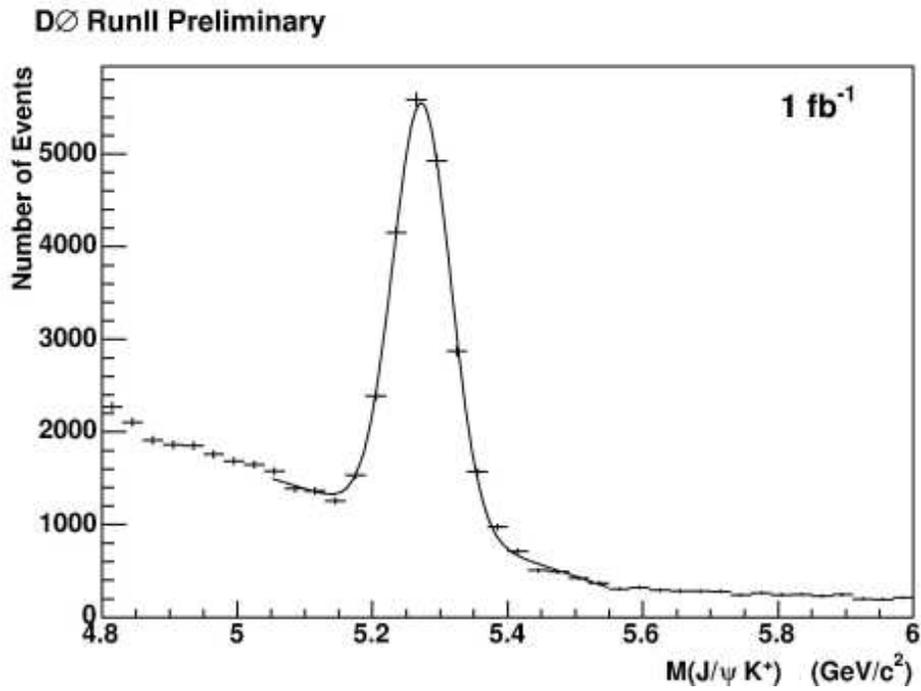}
\caption{The $B^+\to J/\psi K^+$ sample from D0 contains 
16K of signal events and serves as the basis for all 
of their $B^{**}$
measurements.
}
\label{fig:Bstst-jpsik-D0}
\end{figure*}

  The analysis \cite{bstst-D0} 
from the D0 experiment searches for all
three possible decays of the narrow states mentioned 
above. The final state $B^+\pi^-$ is reconstructed. 
The photon coming from $B^{*+}\to B^+\gamma$ decays
is ignored, and leads to a shifted position of the 
mass peak for $B^0_1\to B^{*+}\pi^-$ and 
$B^{*0}_2\to B^{*+}\pi^-$ signal events. The $B^+$
mesons are collected with dimuon trigger in the
channel $B^+\to J/\psi K^+$. The mass distribution
of $B^+$ candidates for the full D0 sample (1~fb$^{-1}$)
is shown in Fig.~\ref{fig:Bstst-jpsik-D0}.

\begin{figure*}[t]
\centering
\includegraphics[width=135mm]{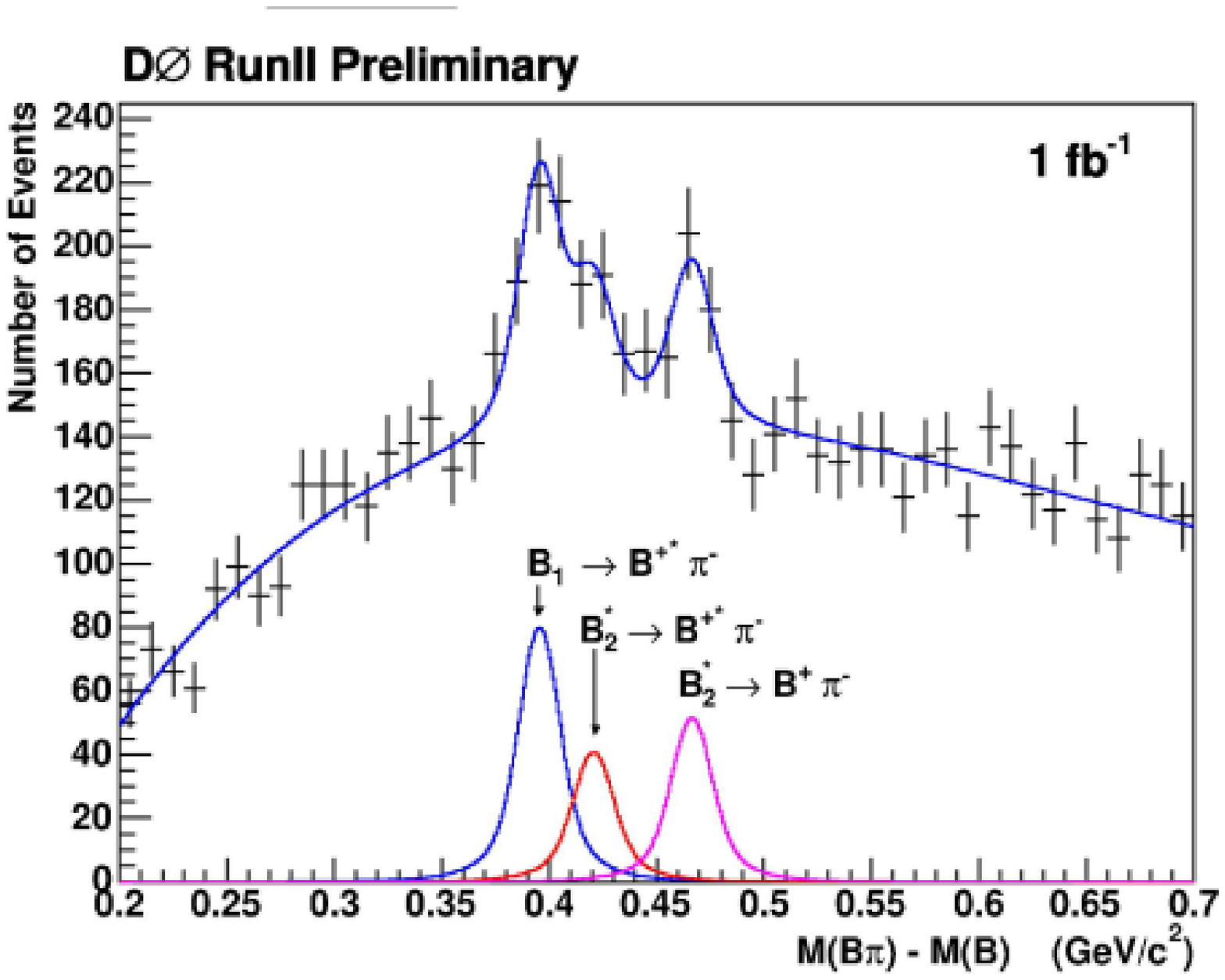}
\caption{Invariant mass difference for the $B^{**}$ 
candidates in the analysis from D0 with fit overlaid.
}
\label{fig:Bstst-dm-D0}
\end{figure*}

   The mass difference $m(B\pi)-m(B)$ for the $B^+\pi^-$ candidates is
shown in Fig.~\ref{fig:Bstst-dm-D0}. 
This is the first observation of separate peaks for
the narrow $B^{**}$ states. D0 proceeds to fit this mass spectrum,
assuming that the widths of the two narrow resonances are the same and
fixing the mass difference between the $B^{*}$ and $B^+$ to
45.78~MeV/$c^2$ \cite{pdg}. The fit returns the masses and the
width of these states:
  \[ M(B_1) = 5720.8\pm 2.5\pm 5.3~MeV/c^2 ~,\]
  \[ M(B^*_2) - M(B_1) = 25.2\pm 3.0\pm 1.1~MeV/c^2 ~,\]
  \[ \Gamma(B_1) = \Gamma(B^*_2) = 6.6\pm 5.3\pm 4.2~MeV/c^2 ~.\]

The D0 also reports the production rates for these resonances:
  \[
     \frac{Br(b\to B^0_J\to B\pi)}
          {Br(b\to B^+)} = 
              0.165\pm 0.024\pm 0.028 ~,
  \]

  \[
     \frac{Br(B^*_2\to B^*\pi)}
          {Br(B^*_2\to B^{(*)}\pi)} = 
              0.513\pm 0.092\pm 0.115 ~,
  \]

  \[
     \frac{Br(B_1\to B^{*+}\pi)}
          {Br(B_J\to B^{(*)}\pi)} = 
              0.545\pm 0.064\pm 0.071 ~.
  \]

\begin{figure*}[t]
\centering
\includegraphics[width=80mm]{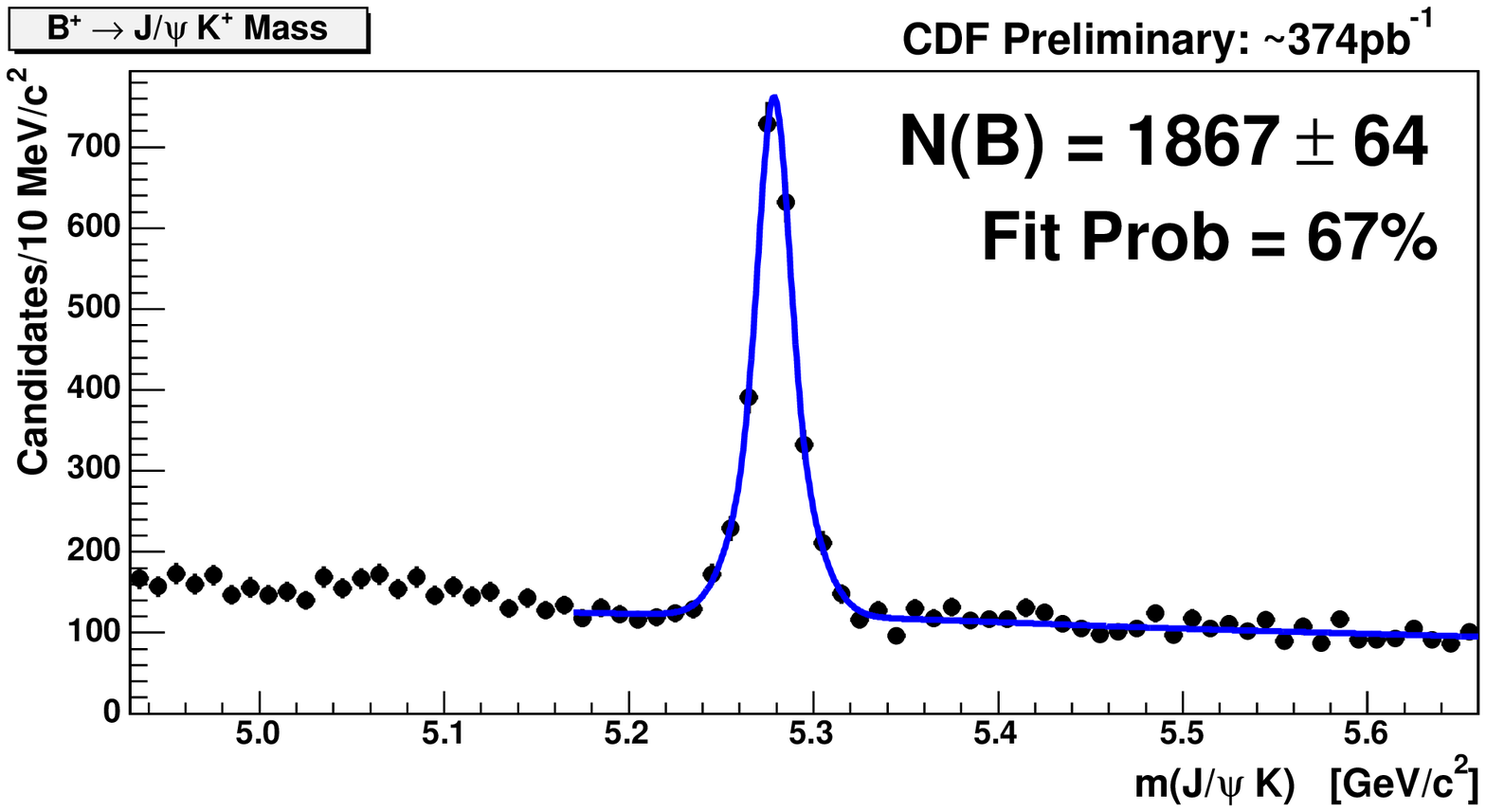}
\includegraphics[width=80mm]{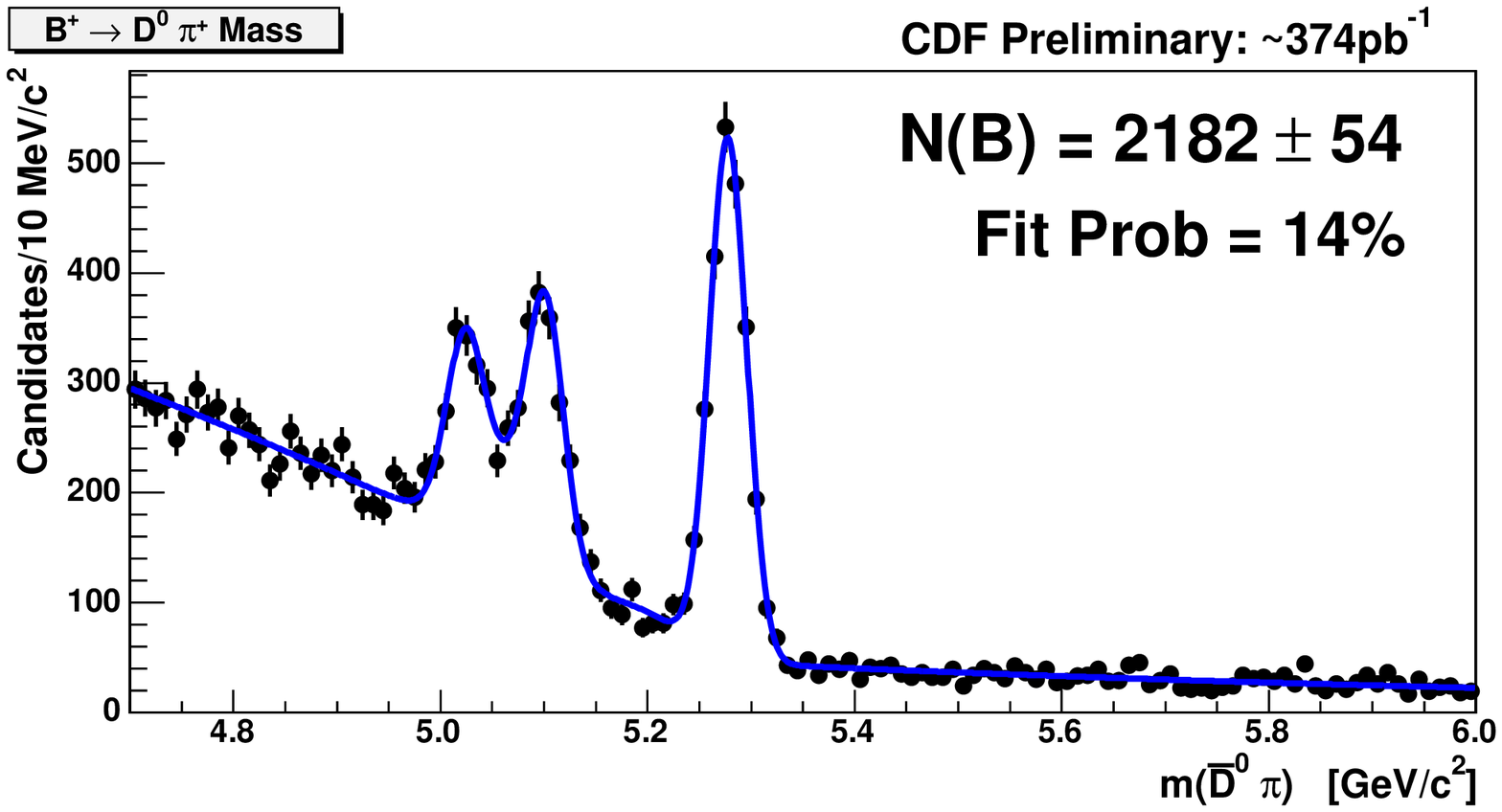}
\caption{The $B^+\to J/\psi K^+$ and $B^+\to D^0\pi^+$
samples from CDF
serve as the basis for all their $B^{**}$
measurements.
}
\label{fig:Bstst-bplus-cdf}
\end{figure*}

\begin{figure*}[t]
\centering
\includegraphics[width=80mm]{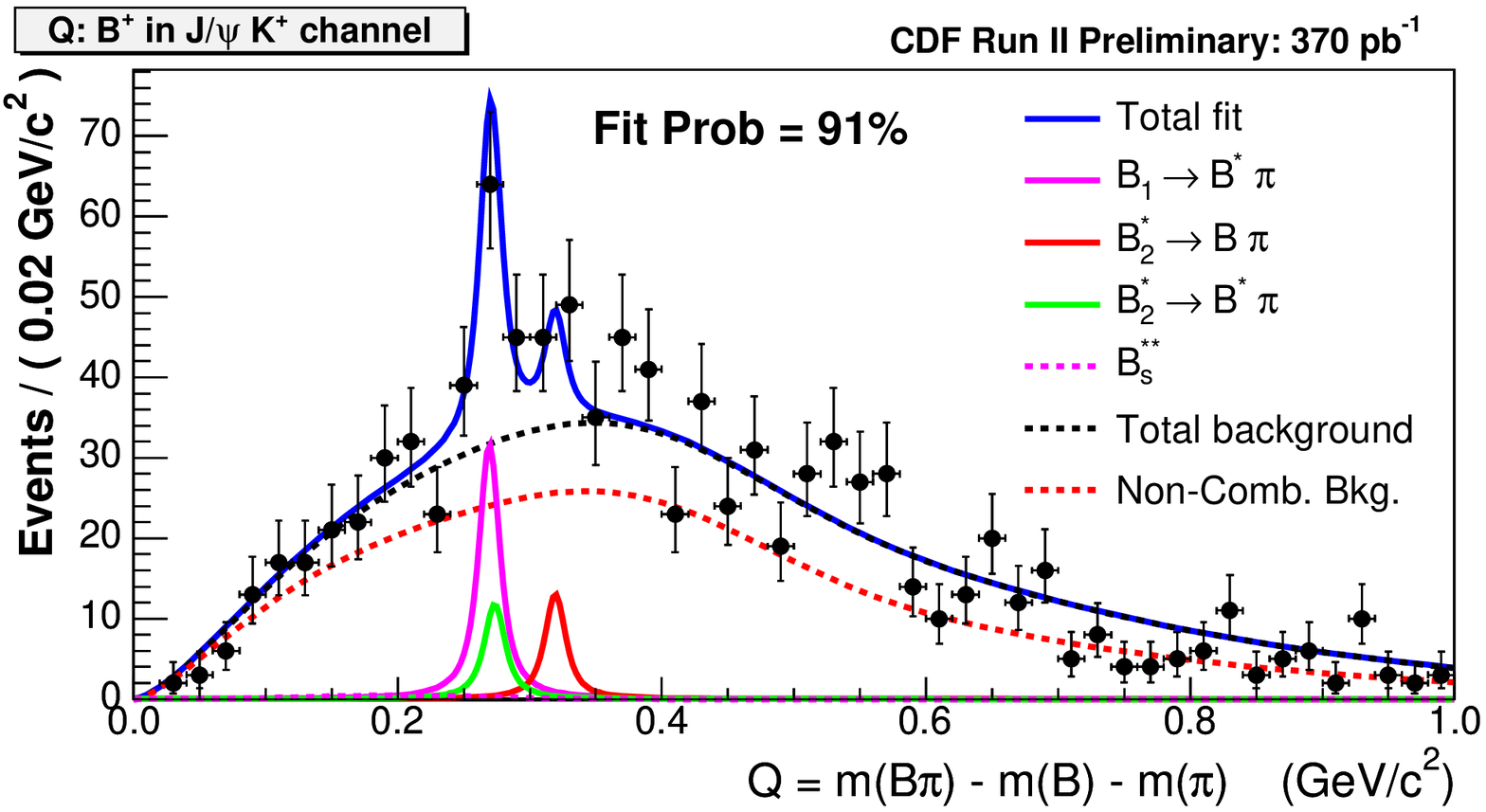}
\includegraphics[width=80mm]{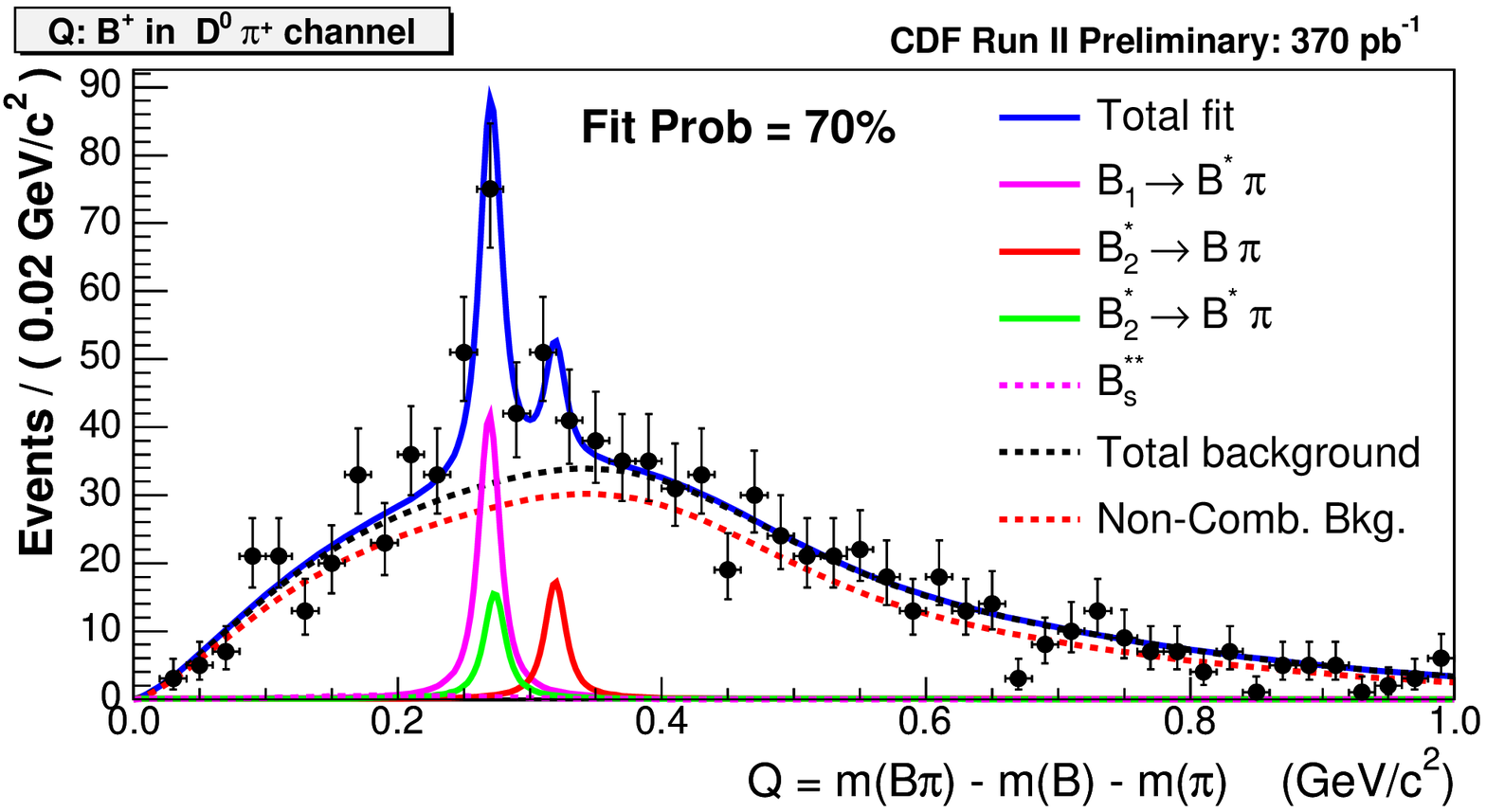}
\caption{Invariant mass difference for the $B^{**}$ 
candidates in the analysis from CDf with fit overlaid.
The two plots correspond to the two $B^+$ samples 
shown in Fig.~\ref{fig:Bstst-bplus-cdf}
}
\label{fig:Bstst-dm-cdf}
\end{figure*}

  The CDF experiment performs a similar analysis. The same
three decays of the $B^0_1$ and $B^{*0}_2$ are the subject
of the measurement. The CDF sample of $B^+$ contains two
signatures: $B^+\to J/\psi K^+$ and $B^+\to D^0\pi^+$
(see Fig.~\ref{fig:Bstst-bplus-cdf}) where the combined yield on 374~pb$^{-1}$
of data is order of 4000 signal candidates. The mass difference
$m(B\pi)-m(B)-m(\pi)$ for the reconstructed $B^{**}$ candidates 
is shown in  Fig.~\ref{fig:Bstst-dm-cdf}. 
The fit of the mass spectrum is 
performed with the widths of both $B^{**}$ fixed
to the theoretical expectation $\Gamma = 16\pm 6~MeV/c^2$
\cite{bstst-width}, and the ratio 
$B^{*0}_2\to B^*\pi/B^{*0}_2\to B\pi$ is assumed to be 
$1.1\pm0.3$ \cite{bstst-ratio}. The result of the fit is
two mass measurements:

  \[ M(B_1)   = 5734\pm 3\pm 2~MeV/c^2 ~,\]
  \[ M(B^*_2) = 5738\pm 5\pm 1~MeV/c^2 ~.\]


\section{Mass measurement of $B^{*0}_{s2}$ }

The heavy-light system $b\bar{s}$ is similar in its
behavior to the $b\bar{u}$ and $b\bar{d}$ systems.
As well, the HQET predicts two narrow and two wide
$B^{**}_s$ states. These are even more difficult to
study because of lower production rates of $B_s$ 
mesons in comparison to more common $B^0$ and $B^+$.

\begin{figure*}[t]
\centering
\includegraphics[width=135mm]{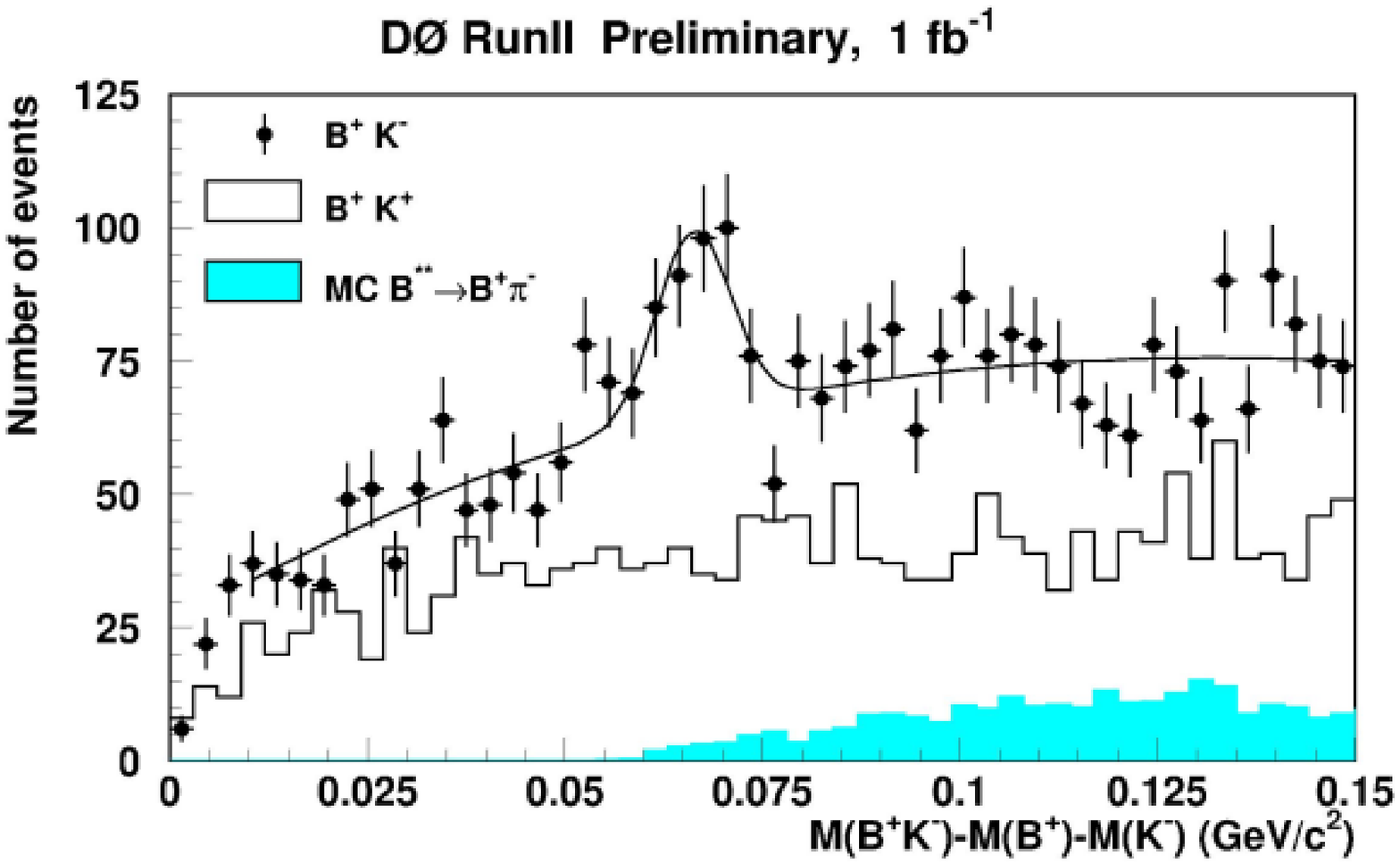}
\caption{Invariant mass difference for the $B^{*0}_{s2}$ 
candidates in the analysis from D0 with fit overlaid.
}
\label{fig:BsStst-dm-D0}
\end{figure*}

  The D0 reports observation of a mass peak that is 
interpreted as $B^{*0}_{s2}$. Due to the isospin
conservation the decays of $B^{**}_s$ to $B_s\pi$
are highly suppressed, thus the D0 team is looking
at the $B^+K^-$ signature. In the analysis, the same
sample of $B^+\to J/\psi K^+$ events, as described in
Sec.~\ref{sec:bstst}, is used. The invariant mass
difference $m(BK)-m(B)-m(K)$ for the $B^{**}_s$
candidates is shown in Fig.~\ref{fig:BsStst-dm-D0}. 
A clear peak is observed
with the significance over $5\sigma$. The D0 asserts that
the observed peak should be identified with the
$B^{*0}_{s2}$ (for longer discussion see \cite{D0-bsstst})
and measures its mass to be
  \[ M(B^{*0}_{s2}) = 5839.1\pm 1.4\pm 1.5~MeV/c^2 ~.\]


\section{Conclusions}

  With 1~fb$^{-1}$ of data, many exciting results on heavy
flavor physics are presently coming from the Tevatron experiments.
In this paper we have seen interesting results on heavy
flavor spectroscopy. CDF narrows the $J^{PC}$ of the X(3872)
down to $1^{++}$ and $2^{-+}$. $B^{**}$ states are now being resolved
and precisely measured. The $B_c$ properties are much
more accessible than in the past. Overall, it is good time
for flavor physics at the Tevatron as we are on the way
to collecting multi-fb$^{-1}$ of data.


\bigskip 
\begin{acknowledgments}
   I would like to thank the experts on the analyses reported
in this paper for provided material and explanations. I also
appreciate help and useful comments from B physics group
conveners from the CDF and D0 experiments. 
\end{acknowledgments}

\bigskip 


\end{document}